\documentclass[preprintnumbers, floatfix,letterpaper,aps,prd,epsfig,nofootinbib,twocolumn]{revtex4-1}
\pdfoutput=1
\usepackage{bm,graphicx,dcolumn,epstopdf,epsf, latexsym,mathbbol, amssymb,amsmath,color,slashed, mathrsfs,mathcomp,simplewick}
\usepackage[center]{subfigure}
\usepackage{multirow}
\usepackage{makecell}
\usepackage[colorlinks,linkcolor=blue,citecolor=blue,urlcolor=blue]{hyperref}
\usepackage[center]{subfigure}
\begin{document}
\allowdisplaybreaks
%%%%%%%%%%%%%%%%%%%%%%%%
 \newcommand{\bq}{\begin{equation}}
 \newcommand{\eq}{\end{equation}}
 \newcommand{\ban}{\begin{align}}
 \newcommand{\ean}{\end{align}}
  \newcommand{\nb}{\nonumber}
 \newcommand{\lb}{\label}
 \newcommand{\f}{\frac}
 \newcommand{\p}{\partial}
%%%%%%%%%%%%%%%%%%%%%%%%%
\newcommand{\PRL}{Phys. Rev. Lett.}
\newcommand{\PLB}{Phys. Lett. B}
\newcommand{\PRD}{Phys. Rev. D}
\newcommand{\CQG}{Class. Quantum Grav.}
\newcommand{\JCAP}{J. Cosmol. Astropart. Phys.}
\newcommand{\JHEP}{J. High. Energy. Phys.}
\newcommand{\NPB}{Nucl. Phys. B}
\newcommand{\Doi}{https://doi.org}
 %%%%%%%%%%%%%%%%%%%%%%%%
\title{Power Spectra of Slow-Roll inflation in the consistent $D\to 4$ Einstein-Gauss-Bonnet gravity}

\author{Tian-Chen Li${}^{a, b}$}
\email{litianchen@zjut.edu.cn}

\author{Tao Zhu${}^{a, b}$}
\email{zhut05@zjut.edu.cn  (Corresponding author)}

\author{Anzhong Wang${}^{c}$}
\email{anzhong\_wang@baylor.edu}

\affiliation{${}^{a}$Institute for Theoretical Physics \& Cosmology, Zhejiang University of Technology, Hangzhou, 310023, China\\
${}^{b}$ United Center for Gravitational Wave Physics (UCGWP),  Zhejiang University of Technology, Hangzhou, 310023, China 
${}^{c}$ GCAP-CASPER, Physics Department, Baylor University, Waco, TX 76798-7316, USA}

\date{\today}

\begin{abstract}

The slow-roll inflation which took place at extremely high energy regimes is in general believed to be sensitive to the high-order curvature corrections to the classical general relativity (GR). In this paper, we study the effects of the high-order curvature term, the Gauss-Bonnet (GB) term, on the primordial scalar and tensor spectra of the slow-roll inflation in the consistent $D \to 4$ Einstein Gauss-Bonnet (4EGB) gravity. The GB term is incorporated into gravitational dynamics via the re-scaling of the GB coupling constant $\alpha \to \alpha/(D-4)$ in the limit $ D\to 4$. For our purpose, we calculate explicitly the primordial scalar and tensor power spectra with GB corrections accurate to the next-to-leading order in the slow-roll approximation in the slow-roll inflation by using the third-order uniform asymptotic approximation method. The corresponding spectral indices and their runnings of the spectral indices for both the scalar and tensor perturbations as well as the ratio between the scalar and tensor spectra are also calculated up to the next-to-leading order in the slow-roll expansions. These results represent the most accurate results obtained so far in the literature.  {In addition, by studying the theoretical predictions of the scalar spectral index and the tensor-to-scalar ratio with Planck 2018 constraint in a model with power-law potential, we show that the second-order corrections are important in future measurements.}

\end{abstract}

%\pacs{98.80.Cq, 98.80.Qc, 04.50.Kd, 04.60.Bc}

\maketitle

\section{Introduction}
\renewcommand{\theequation}{1.\arabic{equation}} \setcounter{equation}{0}

The inflationary theory provides a successful solution to the problems of the standard big bang cosmology for instance the flatness problem and the horizon problem. It also explains successfully the almost scale-invariant and nearly Gaussian spectra of primordial density perturbations \cite{Guth:1980zm, Starobinsky:1980te, Sato:1980yn, Baumann:2009ds}, which eventually evolute to generate the large-scale structure (LSS) observed today in the universe and the cosmic microwave background (CMB) temperature anisotropies, which have been detected with high precision by WMAP \cite{WMAP:2010qai,Larson:2010gs}, PLANCK \cite{Planck:2015sxf, Planck:2015zfm}, and other CMB experiments.

Inflation took place at an extremely high energy regime in the early universe. In this regime, the standard theory of the slow-roll inflation in GR also suffers from several conceptional problems, such as the trans-Planckian problem \cite{Martin:2000xs, Brandenberger:2012aj} and the initial singularity problem \cite{Borde:1993xh, Borde:2001nh}. This is because the classical GR is usually
expected to be broken down at such an extremely high energy regime. Thus, the inflationary theory in GR with some corrections can be regarded as the effective theory of the complete UV quantum gravity. This has led to a large amount of research works for considering possible high-order curvature corrections to slow-roll inflation that arise from radiative corrections of quantum gravity, for example, Horava-Lifshitz gravity \cite{Wang:2017brl} and string/M-theory \cite{Baumann:2014nda}. 

The most important high-order curvature terms are the GB term and its Lovelock generalization, which have been extensively studied in various alternative theories beyond GR. However, because the GB term is a topological invariant in four dimensions, this term can have contributions to the gravitational dynamics only when it is coupled to a matter field. Latterly, Glavan and Lin formulated a new theory with GB term in four dimensions (4EGB) by rescaling the GB coupling constant $\alpha \rightarrow \alpha/(D - 4)$ in the limit $D \rightarrow 4$ \cite{Glavan:2019inb}. With this scaling, it is argued that in the limit $D \rightarrow 4$, the GB term can make nontrivial contributions to the gravitational dynamics. The 4EGB gravity has been extended to higher-order Lovelock gravity in \cite{Hennigar:2020lsl, Fernandes:2020nbq}. {However, there are also certain criticisms of the new 4EGB theory \cite{Ai:2020peo,Gurses:2020ofy,Lu:2020iav,Kobayashi:2020wqy,Hennigar:2020lsl,Fernandes:2020nbq,Shu:2020cjw,Bonifacio:2020vbk,Mahapatra:2020rds}.} Someone argued that this theory lacks an intrinsically four-dimensional description in terms of a covariantly-conserved rank-2 tensor in four dimensions  \cite{Gurses:2020ofy}. The vacua of the model are ill-defined too \cite{Shu:2020cjw}. Several variants of the 4EGB theory are proposed for the purpose of curing these problems, It is shown that by compressing a higher dimensional theory down to a $D-4$ dimensional maximally symmetric space and redefining a few parameters, the 4EGB theory can be reformulated to a specific class of the Horndeski theory \cite{Lu:2020iav}. Its reformulation and Lovelock generalization as a scalar-tensor theory have been deduced \cite{Kobayashi:2020wqy,Lu:2020iav}. In these realizations, the original 4EGB theory is reformulated to a scalar-tensor theory, with a coupling between the scalar field and the GB term. And thus such theories, in general, propagate three degrees of freedom.  {Note that in such a scalar-tensor extension, it is also shown that the scalar field could be strongly coupled in a cosmological background \cite{Kobayashi:2020wqy} and in a flat background \cite{Bonifacio:2020vbk}. }Another way to cure the pathologies of the original 4EGB theory is to break time diffeomorphism invariance but preserve the spatial one \cite{Aoki:2020lig}. It is shown that this spatial covariant 4EGB gravity (i.e. the consistent $D\to 4$ Einstein-Gauss-Bonnet gravity) can only have two degrees of freedom due to a Lagrangian multiplier term in the gravitational action \cite{Aoki:2020lig}.

In this paper, we will especially pay attention to the spatial covariant 4EGB gravity in the early universe, and their corrections to the standard slow-roll inflationary perturbations. It is interesting to note that the GB corrections to slow-roll inflationary spectra in the scalar-Gauss-Bonnet gravity have already been investigated and expanded at length in several works \cite{Jiang:2013gza, Guo:2010jr, Koh:2014bka, Satoh:2010ep,Satoh:2008ck,vandeBruck:2015gjd,Satoh:2007gn, Chakraborty:2018scm}. These papers have studied the primordial perturbation spectra with the GB corrections, and compared it with observed values. It is interesting to note that in the slow-roll inflation, the coupling between the scalar field and the GB term generates a {\em time-dependent sound speed} related to the equation of motion for scalar and tensor perturbations. In the spatial covariant 4EGB theory, it not only predicts time-dependent sound speeds associated with the tensor perturbations in the slow-roll inflation but also modifies the linear dispersion relation of the tensor modes. The primordial power spectra and non-Gaussianities with both the effects of the time-dependent sound speeds and modified dispersion have been explored in \cite{Aoki:2020ila}. However, for calculating the perturbation spectra, the previous works have supposed this {\em time-dependent sound speed} as well as the slow-roll quantities as constants. In fact, this approach will be invalid when it is used to calculate the primordial spectra above the first-order in the slow-roll approximation.

With the emergence of new high-accuracy cosmological data, such as the Planck data, one expects to get tighter constraints on the theory. Thus, it is important to compare the theoretical predictions of specific inflation models with the observational data. With the rapid development of technology, the CMB observations and the incoming large-scale structures surveys will become more and more accurate. For example, the data from the stage-4 CMB experiments and the LiteBird satellite provide will not only increase the sensitivity of primordial gravitational waves ($\epsilon_1$) but also explore much higher multipoles \cite{CMB-S4:2016ple,SimonsObservatory:2018koc,Mallaby-Kay:2021tuk,LiteBIRD:2022cnt, Auclair:2022yxs}. The Euclid satellite and other ground surveys are going to measure the small scales in the matter power spectrum \cite{LSSTScience:2009jmu,Lacasa:2019flz,Euclid:2021qvm, Auclair:2022yxs}. All these measurements will be sensitive to the high-order corrections to the slow-roll perturbation spectra beyond the leading-order in the slow-roll approximations \cite{Adshead:2010mc,Martin:2014rqa,Sprenger:2018tdb}. In this situation, in order to analyze the data with future observations, the theoretical predictions must be more precise and go beyond the leading slow-roll orders \cite{Auclair:2022yxs, Martin1, Martin2}.

With these considerations, it is highly demanded to consider the slow-roll approximation beyond the leading-order in the calculations of the inflationary perturbation spectra. For this purpose, one has to consider both the effects of the time variation of the sound speeds and the modified dispersion relation that arise in the spatial covariant 4EGB theory. Both the time-dependent propagating sound speeds and the modified dispersion relation can make important corrections in the primordial scalar and tensor perturbation spectra. However, considerations of the time variation of the sound speed and the modified dispersion relation make it very difficult to calculate the corresponding power spectra. In this paper, in order to calculate the primordial perturbation spectra and their spectral indices to desired precision, we employ the uniform asymptotic approximation developed in a series of papers \cite{Zhu:2013fha,Zhu:2013upa,Zhu:2014wfa,Zhu:2015ata, Zhu:2016srz,Zhu:2014wda,Zhu:2014aea,Zhu:2015xsa,Zhu:2015owa}. This approximation provides a better treatment to equations with turning points and poles, and has been widely applied in calculating primordial spectra for various inflation models \cite{Zhu:2013fha,Zhu:2013upa,Zhu:2014wfa,Zhu:2015ata, Zhu:2016srz,Zhu:2014wda,Zhu:2014aea,Zhu:2015xsa,Zhu:2015owa} and applications in studying the reheating process \cite{Zhu:2018smk} and quantum mechanics \cite{Li:2019cre}. The main goal of this paper is to use the uniform asymptotic approximation to calculate the inflationary observables of the slow-roll inflation in the spatial covariant 4EGB theory with high accuracy.

The content of the paper is arranged as follows. In Sec. II, we present a brief introduction of spatial covariant 4EGB gravity, and in Sec. III, we consider the cosmological perturbations in a flat FRW background, including the linear scalar perturbations and tensor perturbations. Then we calculate explicitly the power spectra, spectral indices, and runnings of the spectral indices of both scalar and tensor perturbations and the ratio between the scalar and tensor spectra in the slow-roll inflation with the GB correction in Sec. IV. {With the obtained expressions of the scalar spectral index and the tensor-to-scalar ratio, we study their predictions with Planck 2018 constraint in a model for specific power-law potential in Sec. V.} Our main conclusions and outlook are summarized in Sec. VI. We also give the most general formulas for calculating the primordial power spectra in the high-order uniform asymptotic approximations in appendixes A and B.

\section{spatial covariant 4EGB gravity}

\renewcommand{\theequation}{2.\arabic{equation}} \setcounter{equation}{0}

In this section, we will briefly introduce the spatial covariant 4EGB gravity. The details of this theory can be found in \cite{Glavan:2019inb,Aoki:2020lig}. In this theory, the dynamical variables are the shift vector $N^i$, lapse function $N$ and spatial metric $\gamma_{ij}$. We can write the metric of a spacetime in the Arnowitt-Deser-Misner (ADM) form,
\begin{eqnarray}
ds^2 = - N^2 dt^2 + \gamma_{ij} (dx^i+N^i dt) (dx^j + N^j dt).
\end{eqnarray}
The action of this theory is as follows,
\begin{eqnarray} 
\lb{4EGB}
S_{\rm EGB}^{\rm 4D} &=& \int dt d^{3}x  \sqrt{\gamma}N {\cal{L}}^{\rm 4D}_{\rm EGB} 
\end{eqnarray}
here $\gamma$ is the determinate of the spatial metric $\gamma_{ij}$ with $\gamma={\rm det}(\gamma_{ij})$ , and
\begin{eqnarray} \lb{2.5}
{\cal{L}}^{\rm 4D}_{\rm EGB} &=& \frac{M_{\rm Pl}^2}{2} \Bigg[2R- {\cal M}+ \frac{\tilde \alpha}{2} \Big(8 R^2 - 4 R {\cal M} -{\cal M}^2 \nb\\
&& - \frac{8}{3}\left(8 R_{ij}R^{ij} - 4 R_{ij} {\cal M}^{ij} - {\cal M}_{ij} {\cal M}^{ij}\right)\Big) \Bigg],\nb\\
\end{eqnarray}
with
\begin{eqnarray}
{\cal M} _{ij} &\equiv& R_{ij} + {\cal K}^{k}_{k} {\cal K}_{ij} - {\cal K}_{ik} {\cal K}^{k}_{j},\\
{\cal M} &\equiv & {\cal M}^i_i,\\
{\cal K}_{ij} &\equiv & \frac{1}{2 N} \left(\partial_t \gamma_{ij} - 2 D_{(i} N_{j)} - \gamma_{ij} D^2 \lambda_{\rm GF}\right).
\end{eqnarray}
In the above, $M_{\rm Pl}^2 = 1/(8 \pi G)$ is the reduced Planck mass, where $G$ being the Newton gravitational coupling constant. $R$, $R_{ij}$ here are respectively the Ricci scalar and the Ricci tensor refer to the spatial-metric $\gamma_{ij}$. $D_i$ is the covariant derivative and can be compatible with the spatial metric. $\lambda_{\rm GF}$ is similar to the Lagrange multipliers, and it imposes a primary constraint on the theory such that it can propagate only two degrees of freedom \cite{Glavan:2019inb,Aoki:2020lig}. The coupling constant $\tilde \alpha$ is the rescaled GB coupling constant $\tilde \alpha \to \alpha/(D-4)$ with the limit $D \to 4$. 

One important feature of the spatial covariant 4EGB gravity is that it don't admit the full diffeomorphism invariance of the four-dimensional spacetime, but break the temporal diffeomorphism. Then this theory only has the time reparametrization symmetry and the three-dimensional spatial diffeomorphism,
\begin{eqnarray}
t &\to& \tilde t = t - f(t), \\
x^i & \to & \tilde x^i =x^i - \zeta^i (t, x^j).
\end{eqnarray}
where $f(t)$ and $\zeta^i (t, x^j)$ are the the infinitesimal generators of the time reparametrization and the spatial diffeomorphism, respectively. With this gauge transformation, the dynamical variables $N$, $N_i$, and $\gamma_{ij}$ can transform as
\begin{eqnarray}
\delta N &=& \zeta^k D_k N+ \dot N f + N \dot f,\nb\\
\delta N_i &=& N_k D_i \zeta^k + \zeta^k D_k N_i + \gamma_{ik} \dot \zeta^k+ \dot N_i f + N_i \dot f, \nb\\
\delta \gamma_{ij} &=& D_i \zeta_j + D_j \zeta_i + f \dot \gamma_{ij}. \lb{gaugeT}
\end{eqnarray}
Here a dot is the derivative of respect to the time $t$. 

\section{Cosmological perturbations in a flat FRW background}
\renewcommand{\theequation}{3.\arabic{equation}} \setcounter{equation}{0}

In this section, we present a brief introduction of the scalar and tensor perturbations of the slow-roll inflation in the spatial covariant 4EGB gravity.

\subsection{Slow-roll inflation}

For later convenience to introduce the cosmological perturbations, we first consider a flat Friedmann-Robertson-Walker (FRW) background,
\begin{eqnarray}
ds^2&=&-dt^2 +a^2(t) \delta_{ij}dx^{i}dx^{j} \nb\\
&=& a^2(\tau) (- d \tau^2 +\delta_{ij}dx^{i}dx^{j} ), 
\end{eqnarray}
where $a(t)$ is the scalar factor of the universe , $t$ is the cosmic time, and $\tau$ is the conformal time. For studying the inflationary cosmology and cosmological perturbations, we can consider the theory in (\ref{4EGB}) with a scalar field $\phi$ with a potential $V(\phi)$, that is 
\begin{eqnarray}\lb{total}
S = S_{\rm EGB}^{\rm 4D} +S_\phi
\end{eqnarray}
 where
  \begin{eqnarray}
  S_{\phi}=\int dtd^{3} x \sqrt{\gamma} N \left[\frac{1}{2} g^{\mu \nu} \partial_{\mu} \phi \partial_{\nu} \phi+V(\phi)\right].
 \end{eqnarray}
Here the scalar field $\phi$ is canonical and coupled to gravity minimally, $\phi$ is the inflation field with potential $V(\phi)$, and $g^{\mu\nu}$ is the metric of the 4-dimensional spacetime.  Thus the modified Friedmann and the Klein-Gordon equations in the  spatial covariant 4EGB gravity can be written as \cite{Glavan:2019inb,Aoki:2020lig}
\begin{eqnarray}
3 M_{\rm Pl}^{2} (H^{2}+ \tilde \alpha H^{4}) =\frac{1}{2} \dot{\phi}^{2}+V(\phi),\\
- 2 M_{\rm Pl}^2 (1+ 2 \tilde \alpha H^2) \dot H = \dot \phi^2,
\end{eqnarray}
and
\begin{eqnarray}
\ddot \phi+3 H \dot \phi+V_{,\phi}=0 \lb{kg_eq},
\end{eqnarray}
where $H\equiv \dot a/a$ is the Hubble parameter with a dot is the derivative with respect to the cosmic time $t$, and $V_{,\phi}=dV(\phi)/d\phi$. With these equations of the background evolutions,  we will respectively consider the cosmological scalar and tensor perturbations in the following subsections.  

To consider the slow-roll inflation, we also need to impose the following slow-roll conditions
\begin{eqnarray}
\frac{1}{2}\dot \phi^2 \ll V,\;\; |\ddot \phi| \ll |3 H \dot \phi|. \lb{srcon}
\end{eqnarray}
Then we can easily introduce the Hubble flow slow-roll parameters $\epsilon_{n}$. The definitions of the Hubble flow slow-roll parameters are
\begin{eqnarray}
&&\epsilon_{n+1} \equiv \frac{d\ln \epsilon_{n}}{d\ln a}, \;\;\epsilon_1 \equiv - \frac{\dot H}{H^2}.
\end{eqnarray}
The conformal time $\tau$ is defined as
\begin{eqnarray}
\tau(t)=\int^t_{t_\text{end}}\frac{dt'}{a(t')},
\end{eqnarray}
where $t_{\text{end}}$ is the time of the slow-roll inflation ends. We also introduce a new parameter to represent the GB effects and simplify the calculation, which is defined as
\begin{eqnarray}
f_{\rm GB} = 2 \tilde \alpha H^2.
\end{eqnarray}
During the slow-roll inflation, this parameter is also slow-varying , and it characterizes the correction of the GB term in the spatial covariant 4EGB gravity. In the slow-roll approximation, we treat this parameter as a new slow-roll parameter in the slow-roll expansion. {In principle, the value of the paramater $f_{\rm GB}$ is not necessary to be at the order to of $\mathcal{O}(\epsilon)$. However, in orde to simplfy the calcualtions of the power spectra later, we assume that $f_{\rm \rm GB} \sim \mathcal{O}(\epsilon)$. Thus the calculations presented in this paper are valid only with this condition. }

\subsection{Scalar perturbations}

Let us consider the linear scalar perturbations around the flat FRLW spacetime in the spatial covariant 4EGB gravity. It is more convenient to use the conformal time $\tau$ to achieve this purpose, and the background variables $(N, N^i, \gamma_{ij})$ is as follows
\begin{eqnarray}
\hat N = a(\tau), \;\; \hat N^i =0, \;\; \hat \gamma_{ij} = a^2(\tau) \delta_{ij},
\end{eqnarray}
where the quantities with a hat denote the background fields only depend on $\tau$. Under this background, we can introduce the scalar perturbations $(A,B,\psi,E,\delta\phi,\delta\lambda)$ as,
\begin{eqnarray}
N = a(\tau) (1+ A), \;\; N^i = \delta^{ij} \partial_j B, \nb\\
\gamma_{ij} = a^2(\tau) [(1+2 \psi)\delta_{ij} + 2 \partial_i \partial_j E], \nb\\
\phi =  \hat \phi + \delta \phi, \;\;\; \lambda_{\rm GF} = \hat \lambda_{\rm GF} + a^2 \delta \lambda.
\end{eqnarray}
Here $\hat \phi$ and $\hat \lambda_{\rm GF}$ are the background fields of $\phi$ and $\lambda_{\rm GF}$ respectively, and only depends on $\tau$.  With the gauge transformation (\ref{gaugeT}), the scalar perturbations can transform in the form
\begin{eqnarray}
A \to \tilde A &=& A +f' + \mathcal{H} f, \\
B \to \tilde B &=& B + \zeta ', \\
\psi \to \tilde \psi &=& \psi + \mathcal{ H}f,\\
E \to \tilde E &=& E  +  \zeta.
\end{eqnarray}
Here we split $\zeta^i = \zeta^{, i} + \zeta_{\perp}^i$ with $\zeta_{\perp}^i$ being the transverse part of the vector $\zeta^i$. Because the sptaial covariant 4EGB gravity breaks the time diffeomorphism, one can chose a gauge like neither $A=0$ nor $\psi=0$. From gauge transformation has been discussed above, there are only two gauges can be chosen:
\begin{eqnarray}
&& (i) \;\; B=0, \;\; {\rm or} \;\; (ii) \;\; E=0.
\end{eqnarray}
Thus, of the four scalar type metric perturbations $(A, B, \psi, E)$, only one of $(E, B)$ can be eliminated by choosing $\zeta(t, x^i)$ in the gauge transformation. This is different from the theory of full diffeomorphism in 4-dimensional spacetime, where either $A$ or $\psi$ can also be eliminated by choosing $f$ freely. 

Then we consider the comoving curvature perturbation $\cal{R}$, which is defined as
 \bq
 \lb{6.6}
 {\cal{R}}  = -\psi + \frac{\cal{H}}{\hat{\phi'}}\delta\phi.
 \eq
Under the above gauge transformation, it is not difficult
to verify that $\mathcal{R} $ is still gauge-invariant even in the current setup. With this definition and a series of tedious calculations, we can obtain the second order action of the terms of $\mathcal{R}$ \cite{Glavan:2019inb,Aoki:2020lig},
\begin{eqnarray}
\lb{S2}
S^{(2)} = M_{\rm Pl}^2\int d\tau d^2 x a^2 \epsilon_1 (1+f_{\rm GB}) \Big[ {\cal R}'^2 + \partial^2{\cal R}^2\Big].\nb\\
\end{eqnarray}
With variation of this double-dip action with respect to ${\cal R}$, the equation of motion for the comoving curvature perturbation ${\cal R}$ is obtained as follows,
\begin{eqnarray}\lb{eom2}
\mathcal{R}''+ \left(2 \mathcal{H}+ \frac{\epsilon_1'}{\epsilon_1} + \frac{f_{\rm GB}'}{1+f_{\rm GB}} \right) \mathcal{R} ' - \partial^{2} \mathcal{R}=0.
\end{eqnarray}
We can define the mode function $u_{\mathcal{R}}$ in terms of the Fourier modes of curvature perturbations as $u_{\mathcal{R}} = z_{\mathcal{R}} \mathcal{R}_k$. Then the equation of motion (\ref{eom2}) can be transformed into the form
\begin{eqnarray}\lb{1_scalar}
u''_{\mathcal{R}}(\tau)+\left( k^2-\frac{z''_{\mathcal{R}}}{z_{\mathcal{R}}}\right) u_{\mathcal{R}}(\tau)=0.
\end{eqnarray}
Here a prime is the derivative with respect to the conformal time $\tau$. $z_{\mathcal{R}}$ and $\frac{z''_{\mathcal{R}}}{z_{\mathcal{R}}}$ can be written as follows
\begin{eqnarray}\lb{zs}
z_{\mathcal{R}} &=& a \sqrt{2 \epsilon_1 (1+f_{\rm GB})}M_{\rm Pl},
\end{eqnarray}
and
\begin{eqnarray}
\frac{z''_{\mathcal{R}}}{z_{\mathcal{R}}} &=& 2 a^2 H^2 \Bigg[1- \frac{1+4 f_{\rm GB}}{2(1+f_{\rm GB})} \epsilon_1 + \frac{3}{4}\epsilon_2 \nb\\
&&~~~~~~~~ + \frac{(3+2f_{\rm GB})f_{\rm GB}}{2(1+f_{\rm GB})^2} \epsilon_1^2 
 - \frac{1 + 5 f_{\rm GB}}{4(1+f_{\rm GB})} \epsilon_1 \epsilon_2  \nb\\
&&~~~~~~~~ + \frac{\epsilon_2^2}{8} + \frac{1}{4} \epsilon_2 \epsilon_3 \Bigg].
\end{eqnarray}

\subsection{Tensor perturbations}

The cosmological tensor perturbations $h_{ij}$ is defined as,
\begin{eqnarray}
\delta \gamma_{ij} = a^2 h_{ij}, \; \delta N^i =0, \;\; \delta N=0,
\end{eqnarray}
where  $h_{ij}$ denotes the  traceless and transverse metric perturbations, namely,
\begin{eqnarray}
\partial^i h_{ij} =0 = h_i^i.
\end{eqnarray}
With the gauge transformation (\ref{gaugeT}), the tensor perturbations are found to be gauge invariant. The above definition requires us to derive equations of motion for tensor perturbations. To this end, the metric perturbation can be first substituted into Eq. (\ref{total}) and then expanded to the quadratic order of $h_{ij}$. After a series of complicated calculations, one obtains \cite{Glavan:2019inb,Aoki:2020lig},
\begin{eqnarray}
S^{(2)}_{h} &=& \frac{M_{\rm Pl}^2}{8} \int d\tau d^3 x a^2(\tau) \Big[(1+2 \tilde \alpha H^2)  h'_{ij}  h'^{ij} \nb\\
&& \;\;\;\; - \left(1+ 2 \tilde \alpha H^2 + 4 \tilde \alpha \dot H \right) \partial_k h_{ij} \partial^k h^{ij} \nb\\
&&~~~~ - \frac{4 \tilde \alpha}{a^2(\tau)} \partial^2 h_{ij} \partial^2 h^{ij}\Big].
\end{eqnarray}
Then one can take the variation of the quadratic action of$h_{ij}$ to obtain the equations of motion of the tensor perturbations, which gives,
\begin{eqnarray}
&&h_{ij}'' + 2 \mathcal{H} \left( 1+ \frac{2 \alpha \dot H}{1+ 2 \alpha H^2} \right) h_{ij}'  \nb\\
&& - \left( 1 + \frac{4 \alpha \dot H}{1+ 2 \alpha H^2 }\right) \partial^2 h_{ij} + \frac{4\alpha/a^2}{1+ 2 \alpha H^2} \partial^4 h_{ij}=0, \nb\\ 
\end{eqnarray}
where $\mathcal{H}\equiv a'/a$ with a {\emph{prime}} representing the derivative with respect to the conformal time $\tau$.

In order to study the evolution of $h_{ij}$, we expand it to the spatial Fourier harmonics,
\begin{eqnarray}
h_{ij}(\tau, x^i) = \sum_{A={\rm R, L}} \int \frac{d^3 k}{(2\pi)^3}  h_A(\tau, k^i)e^{i k_i x^i} e_{ij}^{A}(k^i),\nb\\
\end{eqnarray}
where $e_{ij}^A$ represent the circular polarization tensors. $e_{ij}^A$ satisfy the following relation
\begin{eqnarray}
\epsilon^{i j k} n_i e_{kl}^A = i \rho_A e^{jA}_{~l},
\end{eqnarray}
where  $\rho_{\rm L} =-1$ and $\rho_{\rm R}=1$. It is worth noting that the propagation equations of these two modes are decoupled, and can be transformed into
\begin{eqnarray}\lb{eom}
h_A'' + \left(2 \mathcal{H}+ \frac{f_{\rm GB}'}{1+f_{\rm GB}} \right)  h_A' + \omega_k^2 h_A=0.
\end{eqnarray}
The quantity $\frac{f_{\rm GB}'}{1+f_{\rm GB}}$ in parentheses gives a description of the correction of the friction term, while $\omega_k^2$ describes the correction of the dispersion relation for the tensor perturbations, which reads
\begin{eqnarray}
\omega_k^2 &\equiv& \left(1 - \frac{8 \alpha \dot H}{M_{\rm Pl}^2 + 4 \alpha H^2} + \frac{4 \tilde \alpha /a^2}{1+2 \tilde \alpha H^2}k^2\right)k^2 \nb\\
&=& \left(1+ \frac{2 f_{\rm GB} \epsilon_1}{1+f_{\rm GB}}+ \frac{2 f_{\rm GB}}{1+f_{\rm GB}} \frac{k^2}{a^2H^2}\right)k^2 \nb\\
&=& c_h^2 k^2 + \frac{2 f_{\rm GB}}{1+f_{\rm GB}} \frac{k^2}{a^2H^2}k^2,
\end{eqnarray}
with
\begin{eqnarray}
c_h^2 \equiv 1+ \frac{2 f_{\rm GB} \epsilon_1}{1+f_{\rm GB}}.
\end{eqnarray}

We then define a new variable ${u_h = \frac{1}{2} z_h h_A}$ and change Eq. (\ref{eom}) to
\begin{eqnarray}\lb{1_tensor}
u_h'' + \left[\omega_k^2 - \frac{z_h''}{z_h} \right] u_h=0,
\end{eqnarray}
where
\begin{eqnarray}\lb{zt}
{z_h \equiv  a\sqrt{1+f_{\rm GB}}M_{\rm{pl}}},
\end{eqnarray}
and
\begin{eqnarray}
\frac{z_h''}{z_h} &=& 2 a^2 H^2 \Bigg[1 -\frac{1+4f_{\rm GB}}{2(1+f_{\rm GB})} \epsilon_1+\frac{(3+2 f_{\rm GB})f_{\rm GB}}{2(1+f_{\rm GB})^2} \epsilon_1^2  \nb\\
&&~~~~~~~~~~~~~~~~~ - \frac{f_{\rm GB}}{2(1+f_{\rm GB})} \epsilon_1 \epsilon_2\Bigg].
\end{eqnarray}

\section{Scalar and Tensor Perturbation Spectra in the uniform asymptotic approximation}
\renewcommand{\theequation}{4.\arabic{equation}}\setcounter{equation}{0}
%%%%%%%%%%%%%%%%%%%%%%%%%%%%%%%
%%%%%%%%%%%%%%%%%%%%%%%%%%%%%%%

In this section, we calculated the scalar and tensor spectra of the slow-roll inflation in the spatial covariant 4EGB gravity. The power spectrum calculation method used in this paper is the uniform asymptotic approximation, and the third-order approximation of it are developed in a series papers \cite{Zhu:2013fha,Zhu:2013upa,Zhu:2016srz,Zhu:2014wfa,Zhu:2014wda,Zhu:2014aea,Zhu:2015xsa,Zhu:2015owa,Zhu:2015ata,Zhu:2018smk}. The general formulas for calculating the primordial power spectra from the slow-roll inflation are described in appendix A and B. The primordial power spectra for both the scalar and tensor perturbations are calculated up to the second-order in the slow-roll approximation in the following subsections. 

\subsection{Scalar Spectrum}

We first consider the scalar perturbations. In order to employ the uniform asymptotic approximation, we need to map the equation of motion of $u_{\cal R}$ in (\ref{1_scalar}) into the standard form in (\ref{standard}). We have
\begin{eqnarray}
\lambda^2 \hat g(y)+q(y)= \frac{\nu_{\cal R}^2}{y^2} -1,
\end{eqnarray}
with
\begin{eqnarray}
\lambda^2 \hat g(y) =\tau^2\frac{z''_{\cal R}}{z_{\cal R}}\frac{1}{y^2}, \\
q(y) = -\frac{1}{4y^2},
\end{eqnarray}
where
\begin{eqnarray}
\nu_{\cal R}^2 &=& \frac{1}{4}+\tau^2\frac{z''_{\cal R}}{z_{\cal R}} \\
&=&\frac{1}{4}+2\tau^2 a^2 H^2 \Bigg[1- \frac{1+4 f_{\rm GB}}{2(1+f_{\rm GB})} \epsilon_1 + \frac{3}{4}\epsilon_2 \nb\\
&& +\frac{(3+2f_{\rm GB})f_{\rm GB}}{2(1+f_{\rm GB})^2} \epsilon_1^2 - \frac{1 + 5 f_{\rm GB}}{4(1+f_{\rm GB})} \epsilon_1 \epsilon_2 \nb\\
&&+ \frac{\epsilon_2^2}{8} + \frac{1}{4} \epsilon_2 \epsilon_3 \Bigg]
\lb{nu}
\end{eqnarray}
It is easy to see that the function $\lambda^2 \hat g(y)$ has a single turning point $y_0(\bar \tau_0)= - k \bar \tau_0$, which related to $\nu_{\cal R}$ as
\begin{eqnarray}
y_0(\bar \tau_0) = - k \bar \tau_0 = \nu_{\cal R}(\bar \tau_0).
\end{eqnarray}
Hereafter we use a bar over the quantities denoting the quantities evaulated at the turning point.

In order to match the accuracy of forthcoming observations, we need to calculate the scalar spectrum and the corresponding spectral indices up to the next-to-leading order (second-order) in the expansions of the slow-roll approximation. For this purpose, we have to consider the time variations of the slow-roll quantities such as $\nu_{\cal R}$. Considering $\nu_{\cal R}$ is slowly varying, it is convenient to expand it up to the second-order in the slow-roll expansions as that presented in Eq.~(\ref{v1}) in appendix B. For $\nu_{\mathcal{R}}$, the expansion coefficients $\bar \nu_{\mathcal{R} 0}$, $\bar \nu_{\mathcal{R} 1}$, and $\bar \nu_{\mathcal{R} 2}$ in (\ref{v1}) can be calculated from Eqs.(\ref{nu}) and (\ref{zs}), we find
\begin{eqnarray}
\bar \nu_{\mathcal{R} 0}& \simeq&\frac{3}{2}+\bar{\epsilon} _1+\frac{\bar{\epsilon} _2}{2}-\bar{f}_{\text{GB}}\bar{\epsilon} _1+\bar{\epsilon} _1^2+\frac{11\bar{\epsilon} _1\bar{\epsilon} _2}{6}+\frac{\bar{\epsilon} _2 \bar{\epsilon} _3}{6},\nb\\
\\
 \bar \nu_{\mathcal{R}1} & \simeq&-\bar{\epsilon} _1 \bar{\epsilon} _2-\frac{\bar{\epsilon} _2 \bar{\epsilon} _3}{2},
 \end{eqnarray}
 and
 \begin{eqnarray}
 \bar \nu_{\mathcal{R}2} \equiv \frac{d^2 \nu_{\mathcal{R}}}{d\ln^2 (-\tau)} &=& \mathcal{O}(\bar \epsilon_i^3).
\end{eqnarray}
Here a letter with an over bar means that the quantity is at the turning point $\bar y_0$.

Then, using the above expansions, the power spectrum for the curvature perturbation $\mathcal{R}$ can be calculated via Eq.(\ref{formula_pw}). After tedious calculations we obtain,
\begin{widetext}
\begin{eqnarray}\lb{PW_t}
\Delta_{\mathcal{R}}^2(k)&=&\frac{181 \bar{H}^2}{72 e^3 M_{\rm{pl}}^2 \pi^2\bar\epsilon_1}\Big[1-\bar{f}_{\text{GB}}+\bar{f}_{\text{GB}}^2+\Big(2\ln{2}-\frac{496}{181}+\frac{630\bar{f}_{\text{GB}}}{181}-4\bar{f}_{\text{GB}}\ln{2}\Big)\bar\epsilon_1+\Big(\frac{67}{181}-\ln2\Big)\Big(\bar{f}_{\text{GB}}-1\Big)\bar\epsilon_2\nb\\
&+&\Big(\frac{293}{181}-\frac{630\ln{2}}{181}+2\ln{2}\Big)\bar\epsilon_1^2-\Big(\frac{11}{362}+\frac{67\ln2}{181}-\frac{\ln^2 2}{2}\Big)\bar\epsilon_2^2\nb\\
&+&\Big(\frac{\pi^2}{12}-\frac{4231}{1629}+\frac{47\ln{2}}{181}+\ln^2 2\Big)\bar\epsilon_1\bar\epsilon_2
+\Big(\frac{\pi^2}{24}-\frac{86}{1629}+\frac{67\ln2}{181}-\frac{\ln^2 2}{2}\Big)\bar\epsilon_2\bar\epsilon_3\Big],
\end{eqnarray}
\end{widetext}
{where $e$ is the natiral constant.} From the scalar spectrum (\ref{PW_t}), we find that the parameter $\bar f_{\rm GB}$ enters into the scalar spectrum at the leading-order in the slow-roll expansion. 

Then with the scalar power spectrum given above, the scalar spectral index is,
\begin{eqnarray}
n_{\mathcal{R}}-1& \simeq&-2\Big(1-\bar{f}_{\text{GB}}\Big)\bar\epsilon_1-\bar\epsilon_2-2\bar\epsilon_1^2\nb\\
&&-\Big(\frac{677}{181}-2\ln{2}\Big)\bar\epsilon_1\bar\epsilon_2
-\Big(\frac{67}{181}-\ln{2}\Big)\bar\epsilon_2\bar\epsilon_3,\nb\\
\end{eqnarray}
and the running of the scalar spectral index is expressed as
\begin{eqnarray}
\alpha_{\mathcal{R}} &\simeq&-2\bar\epsilon_1\bar\epsilon_2-\bar\epsilon_2\bar\epsilon_3-4\bar{f}_{\text{GB}}\bar\epsilon_1^2+2\bar{f}_{\text{GB}}\bar\epsilon_1\bar\epsilon_2-6\bar\epsilon_1^2\bar\epsilon_2\nb\\
&&-\Big(\frac{677}{181}-2\ln2\Big)\bar\epsilon_1\bar\epsilon_2^2-\Big(\frac{858}{181}-2\ln2\Big)\bar\epsilon_1\bar\epsilon_2\bar\epsilon_3\nb\\
&&-\Big(\frac{67}{181}-\ln2\Big)\Big(\bar\epsilon_2\bar\epsilon_3^2+\bar\epsilon_2\bar\epsilon_3\bar\epsilon_4\Big).
\end{eqnarray}
For scalar spectral index, the new effects denoted by the slow-roll parameter $\bar f_{\rm GB}$ appears at the next-to-leading order, while for the running of the index, they only contribute to the third-order of the slow-roll approximation.

\subsection{Tensor Spectrum}

Now we consider the tensor spectrum. First we need to derive the expressions of $ \bar\nu_{h 0}, \bar\nu_{ h 1}, \bar b_{h0},\bar b_{h1}$, and $\bar c_{h 0}$, which are the expansion coefficients presented in Eqs.~(\ref{v1}, \ref{v2}, \ref{v3}). Repeating similar calculations for scalar perturbations, we obtain
\begin{eqnarray}
\bar \nu_{h 0} &\simeq& \frac{3}{2}+\bar \epsilon _1-\bar{f}_{\text{GB}}\bar \epsilon _1+\frac{4}{3}\bar \epsilon _1 \bar \epsilon _2+\bar \epsilon _1^2,~~~~~~~ \\
\bar \nu_{h1}&\equiv & \frac{d\nu_h}{d\ln(-\tau)} \simeq -\bar \epsilon_1 \bar \epsilon_2,
\end{eqnarray}
and
\begin{eqnarray}
\bar b_{h0}&\simeq&2\bar{f}_{\text{GB}}-2\bar{f}_{\text{GB}}^2-4\bar{f}_{\text{GB}}\bar \epsilon _1,~~~~~~~~ \\
\bar b_{h1}&\equiv & \frac{d b_h}{d\ln(-\tau)}
\simeq 4\bar{f}_{\text{GB}}.
\end{eqnarray}
and
\begin{eqnarray}
\bar c_{h0}&\simeq&1+\bar{f}_{\text{GB}}\bar{\epsilon}_1.
\end{eqnarray}

Then,  the power spectrum for the tensor perturbation $h_k$ reads
\begin{eqnarray}
\Delta_{h}^2(k) &\simeq&\frac{181\bar{H}^2}{36{M_{\rm{pl}}^2} e^3\pi^2}\Big[1+\frac{179}{181}\bar{f}_{\text{GB}}-\frac{6695}{181}\bar{f}_{\text{GB}}^2\nb\\
&&-\Big(\frac{496}{181}-2\ln{2}\Big)\bar{\epsilon}_1+\Big(\frac{321}{362}-\frac{913\ln{2}}{181}\Big)\bar{f}_{\text{GB}}\bar{\epsilon}_1\nb\\
&&+\Big(\frac{293}{181}-\frac{630\ln{2}}{181}+2\ln^2 2\Big)\bar{\epsilon}_1^2\nb\\
&&+\Big(\frac{\pi^2}{12}-\frac{4636}{1629}+\frac{496\ln{2}}{181}-\ln^2 2\Big)\bar{\epsilon}_1\bar{\epsilon}_2\Big], 
\end{eqnarray}
 Also the tensor spectral index and its running are given by
\begin{eqnarray}
n_h &\simeq&-2\bar{\epsilon}_1-\frac{358}{181}\bar{f}_{\text{GB}}\bar{\epsilon}_1-2\bar{\epsilon}_1^2\nb\\
&&-\Big(\frac{496}{181}-2\ln{2}\Big)\bar{\epsilon}_1\bar{\epsilon}_2,
\end{eqnarray}
and
\begin{eqnarray}
\alpha_h&\simeq& -2\bar{\epsilon}_1\bar{\epsilon}_2-\frac{716}{181}\bar{f}_{\text{GB}}\bar{\epsilon}_1^2-\frac{358}{181}\bar{f}_{\text{GB}}\bar{\epsilon}_1\bar{\epsilon}_2-6\bar{\epsilon}_1^2\bar{\epsilon}_2\nb\\
&&-\Big(\frac{496}{181}-2\ln2\Big)\Big(\bar{\epsilon}_1\bar{\epsilon}_2^2+\bar{\epsilon}_1\bar{\epsilon}_2\bar{\epsilon}_3\Big).
\end{eqnarray}

\subsection{Expressions at Horizon Crossing}

In the last two subsections, all the results are expressed in terms of quantities that are evaluated at the turning point. However, usually those expressions were expressed in terms of the slow-roll parameters which are evaluated at the time $\tau_\star$ when scalar or tensor perturbation modes cross the horizon, i.e., $a(\tau_\star) H (\tau_\star) = c_{\mathcal{R}}(\tau_\star) k$ for scalar perturbations and $a(\tau_\star) H (\tau_\star) = c_h(\tau_\star) k$ for tensor perturbations. Consider modes with the same wavenumber $k$, it is easy to see that the scalar and tensor modes left the horizon at different times if $c_{\mathcal{R}}(\tau) \neq c_h(\tau)$.  When $c_{\mathcal{R}}(\tau_\star) > c_h(\tau_\star)$, the scalar mode leaves the horizon later than the tensor mode, and for $c_{\mathcal{R}}(\tau_\star) < c_h(\tau_\star)$, the scalar mode leaves the horizon before the tensor one.

As we have two different horizon crossing times, it is reasonable to rewrite all our results in terms of quantities evaluated at the later time, i.e., we should evaluate all expressions at scalar horizon crossing time $a(\tau_\star) H (\tau_\star) = c_{\mathcal{R}}(\tau_\star) k$ for
 $c_\mathcal{R}(\tau_\star) > c_h(\tau_\star)$ and at tensor-mode horizon crossing $a(\tau_\star) H (\tau_\star) = c_h(\tau_\star) k$ for $c_{\mathcal{R}}(\tau_\star) < c_h(\tau_\star)$.
\subsubsection{$c_{\mathcal{R}}(\tau_\star) > c_h(\tau_\star)$}
We shall re-write all the expressions in terms of quantities evaluated at the time when the scalar-mode horizon crossing $c_{\mathcal{R}}(\tau_\star) > c_h(\tau_\star)$. Skipping all the tedious calculations, we find that scalar spectrum can be written in the form
\begin{widetext}
\begin{eqnarray}\lb{PW_h}
\Delta_{\mathcal{R}}^2(k) &=& \frac{181 H_{\star}^2}{72e^3 M_{\rm{pl}}^2\pi^2\epsilon_\star 1}\Big[1-f_{\star {\rm GB}}+f_{\star {\rm GB}}^2+\Big(2\ln{3}-\frac{496}{181}+\frac{630}{181}f_{\star {\rm GB}}-4\ln{3}f_{\star {\rm GB}}\Big)\epsilon_{\star1}\nb\\
&&+\Big(\ln{3}-\frac{67}{181}+\frac{67}{181}f_{\star {\rm GB}}-\ln{3}f_{\star {\rm GB}}\Big)\epsilon_{\star2}+\Big(\frac{517}{543}-\frac{630\ln{3}}{181}+2\ln^2 3\Big)\epsilon_{\star1}^2\nb\\
&&+\Big(\frac{\pi^2}{12}-\frac{3688}{1629}+\frac{47\ln{3}}{181}+\ln^2 3\Big)\epsilon_{\star1}\epsilon_{\star2}+\Big(\frac{329}{1086}-\frac{67\ln{3}}{181}+\frac{\ln^2 3}{2}\Big)\epsilon_{\star2}^2\nb\\
&&+\Big(\frac{\pi^2}{24}-\frac{86}{1629}+\frac{67\ln{3}}{181}-\frac{\ln^2 3}{2}\Big)\epsilon_{\star2}\epsilon_{\star3}\Big].
\end{eqnarray}
\end{widetext}
where the subscript ``$\star$" denotes evaluation at the horizon crossing. For the scalar spectral index, one obtains
\begin{eqnarray}\label{nR1}
n_{\mathcal{R}}-1&\simeq&-2 \epsilon _{\star1}-\epsilon _{\star2}-2 \epsilon _{\star1}^2+2f_{\star {\rm GB}}\epsilon _{\star1}\nb\\
&&-\Big(\frac{677}{181}-2\ln 3\Big)\epsilon _{\star1}\epsilon _{\star2}\nb\\
&&-\Big(\frac{67}{181}-\ln 3\Big)\epsilon _{\star2}\epsilon _{\star3}.
\end{eqnarray}
The running of the scalar spectral index reads
\begin{eqnarray}
\alpha_{\mathcal{R}}&\simeq&-2 \epsilon_{\star1} \epsilon _{\star2}-\epsilon _{\star2} \epsilon _{\star3}-4f_{\star {\rm GB}}\epsilon_{\star1}^2+2f_{\star {\rm GB}}\epsilon_{\star1}\epsilon_{\star2}\nb\\
&&-6 \epsilon _{\star1}^2 \epsilon _{\star2}-\Big(\frac{677}{181}-2\ln3\Big) \epsilon _{\star1} \epsilon _{\star2}^2\nb\\
&&-\Big(\frac{858}{181}-2\ln3\Big) \epsilon _{\star1} \epsilon _{\star2}\epsilon _{\star3}\nb\\
&&-\Big(\frac{67}{181}-\ln3\Big) \epsilon _{\star2} \epsilon _{\star3}^2.
\end{eqnarray}
Similar to the scalar perturbations, now let us turn to consider the tensor perturbations, which yield
\begin{widetext}
\begin{eqnarray}
\Delta_{h}^2(k)&=&\frac{181H_\star^2}{36e^3{M_{\rm{pl}}^2}\pi^2}\Big[1-\Big(\frac{496}{181}-2\ln3\Big)\epsilon _{\star1}+\frac{179}{181}f_{\star {\rm GB}}-\frac{6695}{181}f_{\star {\rm GB}}^2-\Big(\frac{654}{181}+9\ln2-\frac{716}{181}\ln3\Big)f_{\star {\rm GB}}\epsilon _{\star1}\nb\\
&&+\Big(\frac{517}{543}-\frac{630}{181}\ln3+2\ln^2 3\Big)\epsilon _{\star1}^2+\Big(\frac{\pi^2}{12}-\frac{4636}{1629}+\frac{496}{181}\ln3-\ln^2 3\Big)\epsilon _{\star1}\epsilon _{\star2}\Big].
\end{eqnarray}
\end{widetext}
For the tensor spectral index, we find
\begin{eqnarray}
n_h&\simeq&-2 \epsilon _{\star1}-\frac{358}{181}f_{\star {\rm GB}}\epsilon _{\star1}-2 \epsilon _{\star1}^2\nb\\
&&-\Big(\frac{496}{181}-2\ln3\Big)\epsilon _{\star1} \epsilon _{\star2}.
\end{eqnarray}
Then, the running of the tensor spectral index reads
\begin{eqnarray}
\alpha_h&\simeq&-2 \epsilon _{\star1} \epsilon _{\star2}-6\epsilon _{\star1}^2\epsilon _{\star2}+\frac{716}{181}f_{\star {\rm GB}}\epsilon _{\star1}^2-\frac{358}{181}f_{\star {\rm GB}}\epsilon _{\star1}\epsilon _{\star2} \nb\\
&&- \Big(\frac{496}{181}-2\ln3\Big)\epsilon_{\star1}\epsilon _{\star2}^2-\Big(\frac{496}{181}-2\ln3\Big) \epsilon _{\star1} \epsilon _{\star2} \epsilon _{\star3}.\nb\\
\end{eqnarray}
Finally with both scalar and tensor spectra given above, we can evaluate the tensor-to-scalar ratio at the horizon crossing time$(\tau_\star)$, and find that
\begin{widetext}
\begin{eqnarray}\label{r1}
{r}&\simeq& {16 \epsilon_{\star1}}\Bigg[1+\frac{360}{181}f_{\star {\rm GB}}-36f_{\star {\rm GB}}^2+\Big(\frac{67}{181}-\ln3\Big)\epsilon _{\star2}-\Big(\frac{53844}{32761}+9\ln2-\frac{720}{181}\ln3\Big)f_{\star {\rm GB}}\epsilon _{\star1}\nb\\
 &&+\Big(\frac{24120}{32761}-\frac{360}{181}\ln3\Big)f_{\star {\rm GB}}\epsilon _{\star2}-\Big(\frac{32615}{196566}+\frac{67}{181}\ln3-\frac{\ln^2 3}{2}\Big)\epsilon _{\star2}^2+\Big(\frac{42500}{98283}-\ln3\Big)\epsilon _{\star1}\epsilon _{\star2}\nb\\
 &&+\Big(\frac{86}{1629}-\frac{\pi^2}{24}-\frac{67}{181}\ln3+\frac{\ln^2 3}{2}\Big)\epsilon _{\star2}\epsilon _{\star3}\Bigg].
\end{eqnarray}
\end{widetext}
\subsubsection{$c_{\mathcal{R}}(\tau_\star) < c_h(\tau_\star)$}
For $c_{\mathcal{R}}(\tau_\star) < c_h(\tau_\star)$, as the scalar mode leaves the horizon before the tensor mode does,we shall rewrite all the expressions in terms of quantities evaluated at the time when the tensor mode leaves the Hubble horizon $a(\tau_\star) H (\tau_\star) = c_h(\tau_\star) k$. Skipping all the tedious calculations, we find that the scalar spectrum can be written in the form
\begin{widetext}
\begin{eqnarray}\lb{PS}
\Delta_{\mathcal{R}}^2(k) &=& \frac{181 H_{\star}^2}{72e^3 M_{\rm{pl}}^2\pi^2\epsilon _{\star1}}\Big[1-f_{\star {\rm GB}}+f_{\star {\rm GB}}^2-\Big(\frac{496}{181}-2\ln3\Big)\epsilon _{\star1}-\Big(\frac{67}{181}-\ln3\Big)\epsilon _{\star2}+\Big(\frac{630}{181}-4\ln3\Big)f_{\star {\rm GB}}\epsilon _{\star1}\nb\\
&&+\Big(\frac{67}{181}-\ln3\Big)f_{\star {\rm GB}}\epsilon _{\star2}+\Big(\frac{517}{543}-\frac{630}{181}\ln3+2\ln^2 3\Big)\epsilon _{\star1}^2+\Big(\frac{\pi^2}{12}-\frac{4774}{1629}+\frac{47}{181}\ln3+\ln^2 3\Big)\epsilon _{\star1}\epsilon _{\star2}\nb\\
&&-\Big(\frac{11}{362}+\frac{67}{181}\ln3-\frac{\ln^2 3}{2}\Big)\epsilon _{\star2}^2+\Big(\frac{\pi^2}{24}-\frac{86}{1629}+\frac{67}{181}\ln3-\frac{\l^2 3}{2}\Big)\epsilon _{\star2}\epsilon _{\star3}\Big].
\end{eqnarray}
\end{widetext}
For the scalar spectral index, one obtains
\begin{eqnarray}
n_{\mathcal{R}}-1&\simeq&-2 \epsilon _{\star1}-\epsilon _{\star2}-2 \epsilon _{\star1}^2+2f_{\star {\rm GB}}\epsilon _{\star1}\nb\\
&&-\Big(\frac{677}{181}-2\ln 3\Big)\epsilon _{\star1}\epsilon _{\star2}\nb\\
&&-\Big(\frac{67}{181}-\ln 3\Big)\epsilon _{\star2}\epsilon _{\star3}.
\end{eqnarray}
The running of the scalar spectral index reads
\begin{eqnarray}
\alpha_{\mathcal{R}}&\simeq&-2 \epsilon_{\star1} \epsilon _{\star2}-\epsilon _{\star2} \epsilon _{\star3}-4f_{\star {\rm GB}}\epsilon_{\star1}^2+2f_{\star {\rm GB}}\epsilon_{\star1}\epsilon_{\star2}\nb\\
&&-6 \epsilon _{\star1}^2 \epsilon _{\star2}-\Big(\frac{677}{181}-2\ln3\Big) \epsilon _{\star1} \epsilon _{\star2}^2\nb\\
&&-\Big(\frac{858}{181}-2\ln3\Big) \epsilon _{\star1} \epsilon _{\star2}\epsilon _{\star3}\nb\\
&&-\Big(\frac{67}{181}-\ln3\Big) \epsilon _{\star2} \epsilon _{\star3}^2.
\end{eqnarray}
Similar to the scalar perturbations, now let us turn to consider the tensor perturbations, which yield
\begin{widetext}
\begin{eqnarray}
\Delta_{h}^2(k)&=&\frac{181H_\star^2}{36e^3 {M_{\rm{pl}}^2} \pi^2}\Big[1-\Big(\frac{496}{181}-2\ln3\Big)\epsilon _{\star1}+\frac{179}{181}f_{\star {\rm GB}}-\frac{6695}{181}f_{\star {\rm GB}}^2-\Big(\frac{654}{181}+9\ln2-\frac{716}{181}\ln3\Big)f_{\star {\rm GB}}\epsilon _{\star1}\nb\\
&&+\Big(\frac{517}{543}-\frac{630}{181}\ln3+2\ln^2 3\Big)\epsilon _{\star1}^2+\Big(\frac{\pi^2}{12}-\frac{4636}{1629}+\frac{496}{181}\ln3-\ln^2 3\Big)\epsilon _{\star1}\epsilon _{\star2}\Big].
\end{eqnarray}
\end{widetext}
For the tensor spectral index, we find
\begin{eqnarray}
n_h&\simeq&-2 \epsilon _{\star1}-\frac{358}{181}f_{\star {\rm GB}}\epsilon _{\star1}-2 \epsilon _{\star1}^2\nb\\
&&-\Big(\frac{496}{181}-2\ln3\Big)\epsilon _{\star1} \epsilon _{\star2}.
\end{eqnarray}
Then, the running of the tensor spectral index reads
\begin{eqnarray}
\alpha_h&\simeq&-2 \epsilon _{\star1} \epsilon _{\star2}-6\epsilon _{\star1}^2\epsilon _{\star2}+\frac{716}{181}f_{\star {\rm GB}}\epsilon _{\star1}^2-\frac{358}{181}f_{\star {\rm GB}}\epsilon _{\star1}\epsilon _{\star2} \nb\\
&&- \Big(\frac{496}{181}-2\ln3\Big)\epsilon _{\star1}\epsilon _{\star2}^2-\Big(\frac{496}{181}-2\ln3\Big) \epsilon _{\star1} \epsilon _{\star2} \epsilon _{\star3}.\nb\\
\end{eqnarray}
Finally with both scalar and tensor spectra given above, we can evaluate the tensor-to-scalar ratio at the horizon crossing time $(\tau_\star)$, and find that
\begin{widetext}
\begin{eqnarray}
{r}&\simeq& {16\epsilon _{\star1}}\Bigg[1+\frac{360}{181}f_{\star {\rm GB}}-36f_{\star {\rm GB}}^2+\Big(\frac{67}{181}-\ln3\Big)\epsilon _{\star2}-\Big(\frac{53844}{32761}+9\ln2-\frac{720}{181}\ln3\Big)f_{\star {\rm GB}}\epsilon _{\star1}\nb\\
&&+\Big(\frac{24120}{32761}-\frac{360}{181}\ln3\Big)f_{\star {\rm GB}}\epsilon _{\star2}+\Big(\frac{108022}{98283}-\ln3\Big)\epsilon _{\star1}\epsilon _{\star2}+\Big(\frac{10969}{65522}-\frac{67}{181}\ln3+\frac{\ln^2 3}{2}\Big)\epsilon _{\star2}^2\nb\\
&&+\Big(\frac{86}{1629}-\frac{\pi^2}{24}-\frac{67}{181}\ln3+\frac{\ln^2 3}{2}\Big)\epsilon _{\star2}\epsilon _{\star3}\Bigg].
\end{eqnarray}
\end{widetext}

%\color{red}

\section{A Special Model with Power-law potential}
\renewcommand{\theequation}{5.\arabic{equation}}\setcounter{equation}{0}

With the accuracy of future observations becoming more and more precise, slow-rolling inflation theory requires to be accurate to the higher order. And this is the most important reason why we need to calculate the above results in the second order. In order to show the difference between the second-order corrections and the first-order results, we can use a specific model with a power-law potential. In this section, we study the predicted results of the relationship between the scalar spectral index $n_{\cal R}$ and the tensor-to-scalar ratio $r$ in the first order and the second order in this specific model.

Firstly, the correspondence between the Hubble flow parameters and the potential slow-roll parameters $(\epsilon_V, \eta_V, \xi_V^2)$ can be written as 
\begin{eqnarray}
\epsilon_V& \equiv &\frac{V_{,\phi}^2M_{\rm{pl}}^2}{2V^2} \nb\\
&=&\epsilon_1\frac{\Big[1+f_{\rm{GB}}-\frac{\epsilon_1(1+f_{\rm{GB}})}{3}+\frac{\epsilon_2(1+f_{\rm{GB}})}{6}\Big]^2}{(1+f_{\rm{GB}})\Big[1+\frac{f_{\rm{GB}}}{2}-\frac{\epsilon_1(1+f_{\rm{GB}})}{3}\Big]^2},\nb\\
\end{eqnarray}
\begin{eqnarray}
\eta_V&\equiv&\frac{V_{,\phi\phi}M_{\rm{pl}}^2}{V} \nb\\
&=&\frac{1}{(1+f_{\rm{GB}})^2(1+\frac{f_{\rm{GB}}}{2}-\frac{\epsilon_1(1+f_{\rm{GB}})}{3})}\nb\\
&&\;\; \times \Bigg[2\epsilon_1(1+\frac{5f_{\rm{GB}}}{2}+\frac{3f_{\rm{GB}}^2}{2})-\frac{\epsilon_2}{2}(1+2f_{\rm{GB}}+f_{\rm{GB}}^2)\nb\\
&&\;\;\;\; -\frac{2\epsilon_1^2}{3}(1+\frac{9f_{\rm{GB}}}{2}+3f_{\rm{GB}}^2)-\frac{\epsilon_2^2}{12}(1+2f_{\rm{GB}}+f_{\rm{GB}}^2)\nb\\
&&\;\;\;\; +\frac{5\epsilon_1\epsilon_2}{6}(1+\frac{14f_{\rm{GB}}}{5}+\frac{9f_{\rm{GB}}^2}{5})\nb\\
&&\;\;\;\; -\frac{\epsilon_2\epsilon_3}{6}(1+2f_{\rm{GB}}+f_{\rm{GB}}^2)\Bigg],\nb\\
\end{eqnarray}
and
\begin{eqnarray}
\xi_V&\equiv&\frac{V_{,\phi}V_{,\phi\phi\phi}M_{\rm{pl}}^4}{V^2} \nb\\
&=&\frac{1+f_{\rm{GB}}-\frac{\epsilon_1(1+2f_{\rm{GB}})}{3}+\frac{\epsilon_2(1+f_{\rm{GB}})}{6}}{(1+f_{\rm{GB}})^4(1+\frac{f_{\rm{GB}}}{2}-\frac{\epsilon_1(1+f_{\rm{GB}})}{3})^2}\nb\\
&& \times \Bigg[4\epsilon_1^2(1+4f_{\rm{GB}}+\frac{9f_{\rm{GB}}^2}{2}+\frac{3f_{\rm{GB}}^3}{2})\nb\\
&&-3\epsilon_1\epsilon_2\Big(1+\frac{10f_{\rm{GB}}}{3}+\frac{11f_{\rm{GB}}^2}{3}+\frac{4f_{\rm{GB}}^3}{3}\Big)\nb\\
&&+\frac{\epsilon_2\epsilon_3}{2}(1+3f_{\rm{GB}}+3f_{\rm{GB}}^2+f_{\rm{GB}}^3)\nb\\
&&-\epsilon_1\epsilon_2^2\Big(1+\frac{11f_{\rm{GB}}}{3}+\frac{13f_{\rm{GB}}^2}{3}+\frac{5f_{\rm{GB}}^3}{3}\Big)\nb\\
&&+3\epsilon_1^2\epsilon_2\Big(1+5f_{\rm{GB}}+\frac{19f_{\rm{GB}}^2}{3}+\frac{7f_{\rm{GB}}^3}{3}\Big)\nb\\
&&-\frac{4\epsilon_1^3}{3}(1+8f_{\rm{GB}}+9f_{\rm{GB}}^2+3f_{\rm{GB}}^3)\nb\\
&&-\frac{7\epsilon_1\epsilon_2\epsilon_3}{6}\Big(1+\frac{25f_{\rm{GB}}}{7}+\frac{29f_{\rm{GB}}^2}{7}+\frac{11f_{\rm{GB}}^3}{7}\Big)\nb\\
&&+\frac{1}{6}(\epsilon_2^2\epsilon_3+\epsilon_2\epsilon_3^2+\epsilon_2\epsilon_3\epsilon_4)(1+3f_{\rm{GB}}+3f_{\rm{GB}}^2+f_{\rm{GB}}^3)\Bigg].\nb\\
&&
%\xi_V&=&\frac{V_{,\phi}V_{,\phi\phi\phi}M_{\rm{pl}}^4}{V^2}=\frac{1+f_{\rm{GB}}-\frac{\epsilon_1(1+2f_{\rm{GB}})}{3}+\frac{\epsilon_2(1+f_{\rm{GB}})}{6}}{(1+f_{\rm{GB}})^4(1+\frac{f_{\rm{GB}}}{2}-\frac{\epsilon_1(1+f_{\rm{GB}})}{3})^2}\nb\\
%&&\Big[4\epsilon_1^2(1+4f_{\rm{GB}}+\frac{9f_{\rm{GB}}^2}{2}+\frac{3f_{\rm{GB}}^3}{2})-3\epsilon_1\epsilon_2()\Big]
\end{eqnarray}
Then it is easy to express the Hubble flow parameters in terms of the potential $V(\phi)$ and the 4EGB coupling $f_{\rm{GB}}$ with slow-roll approximation as
\begin{eqnarray}
\epsilon_1&\simeq&\frac{M_{\rm{pl}}^2V_{,\phi}^2}{6V^4}\Big(3V^2-2M_{\rm{pl}}^2V_{,\phi}^2+2M_{\rm{pl}}^2VV_{,\phi\phi}\Big),\nb\\
\epsilon_2&\simeq&-\frac{M_{\rm{pl}}^2}{3V^4}\Big[6M_{\rm{pl}}^2V_{,\phi}^4+3(2+f_{\rm{GB}})V^3V_{,\phi\phi}\nb\\
&&-3(2+f_{\rm{GB}})V^2V_{,\phi}^2+2M_{\rm{pl}}^2V^2V_{,\phi\phi}^2\nb\\
&&+2M_{\rm{pl}}^2V^2V_{,\phi}V_{,\phi\phi\phi}-10M_{\rm{pl}}^2VV_{,\phi}^2V_{,\phi\phi}\Big],\nb\\
\epsilon_3&\simeq&\Big[-6M_{\rm{pl}}^2V_{,\phi}(2V_{,\phi}^3-3VV_{,\phi}V_{,\phi\phi}+V^2V_{,\phi\phi\phi})\Big]\nb\\
&&\Big[6M_{\rm{pl}}^2V_{,\phi}^4+3(2+f_{\rm{GB}})V^3V_{,\phi\phi}-3(2+f_{\rm{GB}})V^2V_{,\phi}^2\nb\\
&&+2M_{\rm{pl}}^2V^2V_{,\phi\phi}^2+2M_{\rm{pl}}^2V^2V_{,\phi}V_{,\phi\phi\phi}\nb\\
&&-10M_{\rm{pl}}^2VV_{,\phi}^2V_{,\phi\phi}\Big]^{-1}.
\end{eqnarray}
With the slow-roll conditions, the number of e-folds can be changed as
\begin{eqnarray}\label{N}
N=\int_{t}^{t_{\rm{end}}}Hdt\simeq\int_{\phi_{\rm{end}}}^{\phi}\frac{V(2-f_{\rm{GB}})}{2M_{\rm{pl}}^2V_{,\phi}}d\phi,
\end{eqnarray}
where $\phi_{\rm{end}}$ can get from by $\epsilon_1(\phi_{\rm{end}})=1$, which means the value at the end of inflation.

To proceed, it is necessary to consider a specific model with a power-law potential,
\begin{eqnarray}
V(\phi)&=&V_0\phi^n,\nb\\
f_{\rm{GB}}&\simeq&\frac{2\alpha V(\phi)}{3M_{\rm{pl}}^2}=\frac{2\alpha V_0\phi^n}{3M_{\rm{pl}}^2},
\end{eqnarray}
where the value of $V_0$ can be determined by Planck 2018 data with a specific value of $n$. Then the Hubble flow can be expressed as
\begin{eqnarray}
\epsilon_1&\simeq&\frac{M_{\rm{pl}}^2n^2(-2M_{\rm{pl}}^2n+3\phi^2)}{6\phi^{4}},\nb\\
\epsilon_2&\simeq&\frac{M_{\rm{pl}}^2n[-2M_{\rm{pl}}^2n+(2+f_{\rm{GB}})\phi^2]}{\phi^4},\nb\\
\epsilon_3&\simeq&\frac{4M_{\rm{pl}}^2n}{-2M_{\rm{pl}}^2n+(2+f_{\rm{GB}})\phi^2}.
\end{eqnarray}
The scalar field at the end of the inflation $\phi_{\rm{end}}$ can be get from
\begin{eqnarray}
\phi_{\rm{end}}^2=\frac{M_{\rm{pl}}^2n^2}{2}.
\end{eqnarray}
Then we can easily get the expression of the e-fold number with Eq.(\ref{N}),
\begin{eqnarray}
N=\frac{\phi^2}{2M_{\rm{pl}}^2 n}-\frac{\phi^2 f_{\rm{GB}}}{2M_{\rm{pl}}^2n(n+2)}-\frac{n}{4}+\frac{n}{4(n+2)}f_{\rm{GB}}^{\rm{end}}.\nb\\
\end{eqnarray}
where $f_{\rm{GB}}^{\rm{end}}$ the value of $f_{\rm{GB}}$ at the end of the inflation, and when $N$ and $n$ are fixed values, $f_{\rm{GB}}^{\rm{end}}$ can be treated as constant.

\begin{figure}
\includegraphics[scale=0.55]{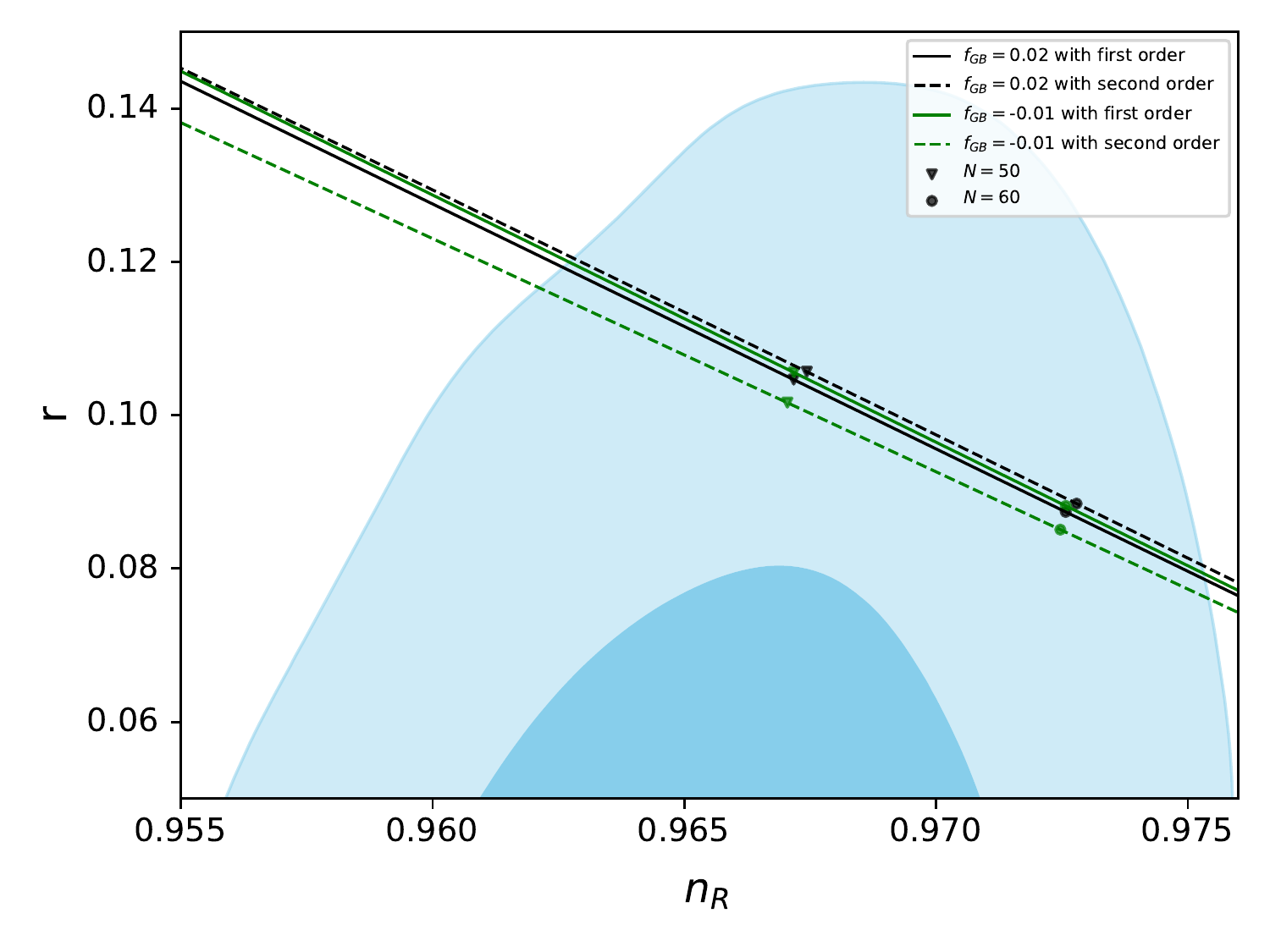}
\caption{{Marginalized constraint (68\% and 95\% confidence level) from $(n_{\cal R}, r)$ using Planck 2018 data, compared to the theoretical predictions of the model with $n=4/3$. For both $f_{\rm{GB}}=0.02$ and $f_{\rm{GB}}=-0.01$, the solid lines give the results with the first-order corrections while the dash gives the predictions with the second-order corrections, respectively.}}
\label{fig1}
\end{figure}

In the case of $c_{\cal R}(\eta_{\star})>c_h(\eta_{\star})$, substituting the above results into Eqs.~(\ref{nR1}) and (\ref{r1}), it is easy to find that $n_{\cal R}$ depends on the parameter $f_{\rm{GB}}$, which is different from the first-order results. For the tensor-to-scalar ratio $r$, it is directly related to both $n$ and $f_{\rm{GB}}$. In Fig.~\ref{fig1}, we plot the theoretical predictions of $n_{\cal R}-r$ relation with $n=4/3$ for different values of $f_{\rm{GB}}$ in comparison with the observational data. The contours are taken from the Planck 2018 TT+TE+EE+lowP data \cite{Planck:2018vyg}. For different values of $f_{\rm{GB}}$, we respectively plot them with the first and the second corrections. 
%It is clear to find that, the difference of $n_{\cal R}-r$ relation between different $f_{\rm{GB}}$ on second-order is larger than ones on first-order.

Considering about the smaller errors on $n_R$ and $r$ in the future, the second-order corrections to both $n_R$ and $r$ in the slow-roll approximation are important. In the forthcoming experiments, especially the Stage IV experiments \cite{CMB-S4:2016ple,SimonsObservatory:2018koc,Mallaby-Kay:2021tuk,LiteBIRD:2022cnt}, the errors of the measurements on both $n_R$ and $r$ will be smaller than $10^{-3}$, namely $\sigma(n_{\cal R}, r)<10^{-3}$. That means it is necessary to calculate the second-order corrections of $n_{\cal R}$ and $r$ to the magnitude of $\mathcal{O}(10^{-3})$ for getting more accurate constraints on the parameters of the models. On this order of magnitude, the first-order and second-order results will be easily distinguished by future experiments. It is worth noting that the contributions of second-order corrections are affected by the parameters of the models. 

\begin{figure}[!t]
\subfigure[]{
\begin{minipage}[t]{1\linewidth}
\includegraphics[width=8cm]{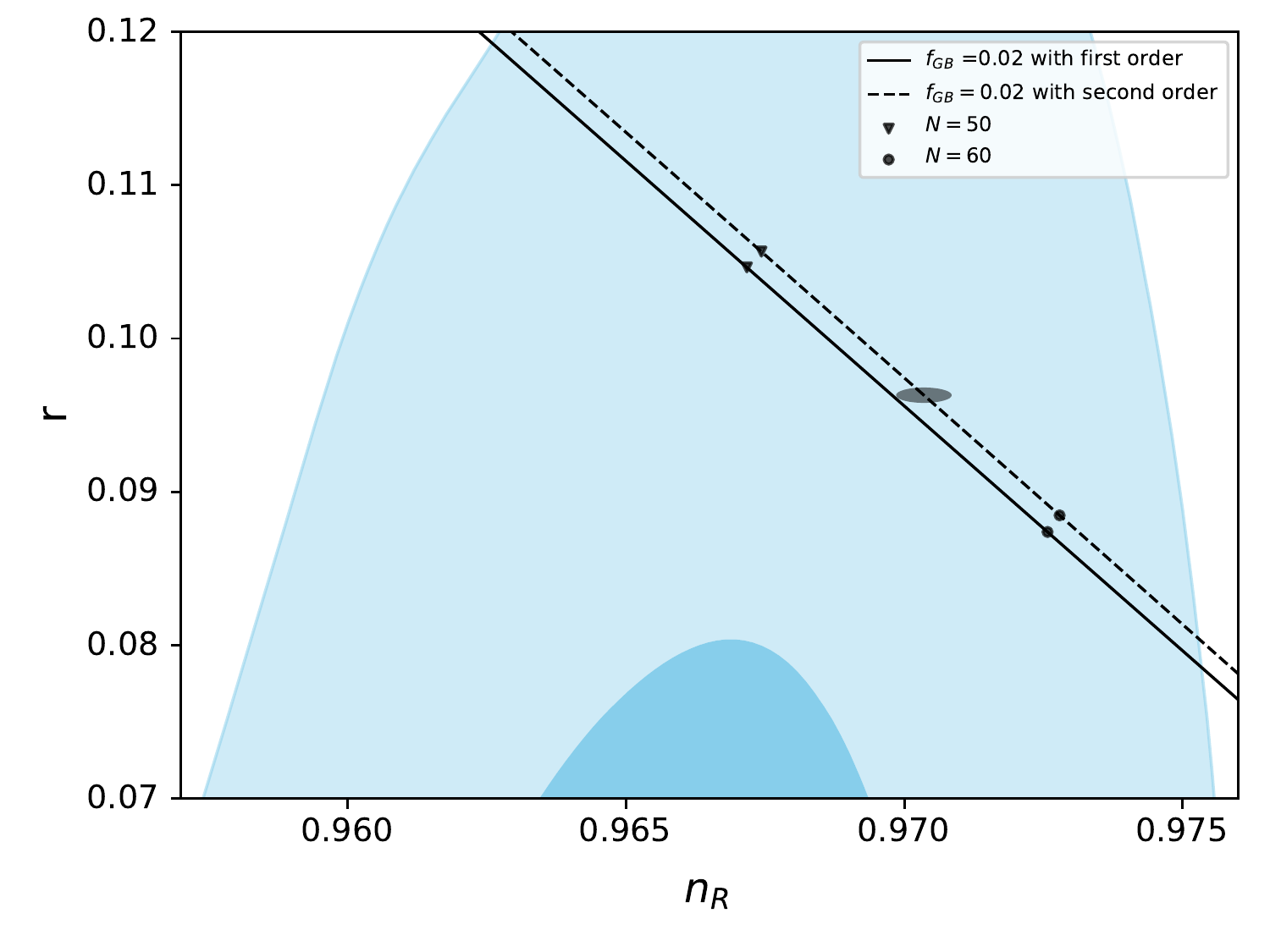}
\end{minipage}
}
\subfigure[]{
\begin{minipage}[t]{1\linewidth}
\includegraphics[width=8cm]{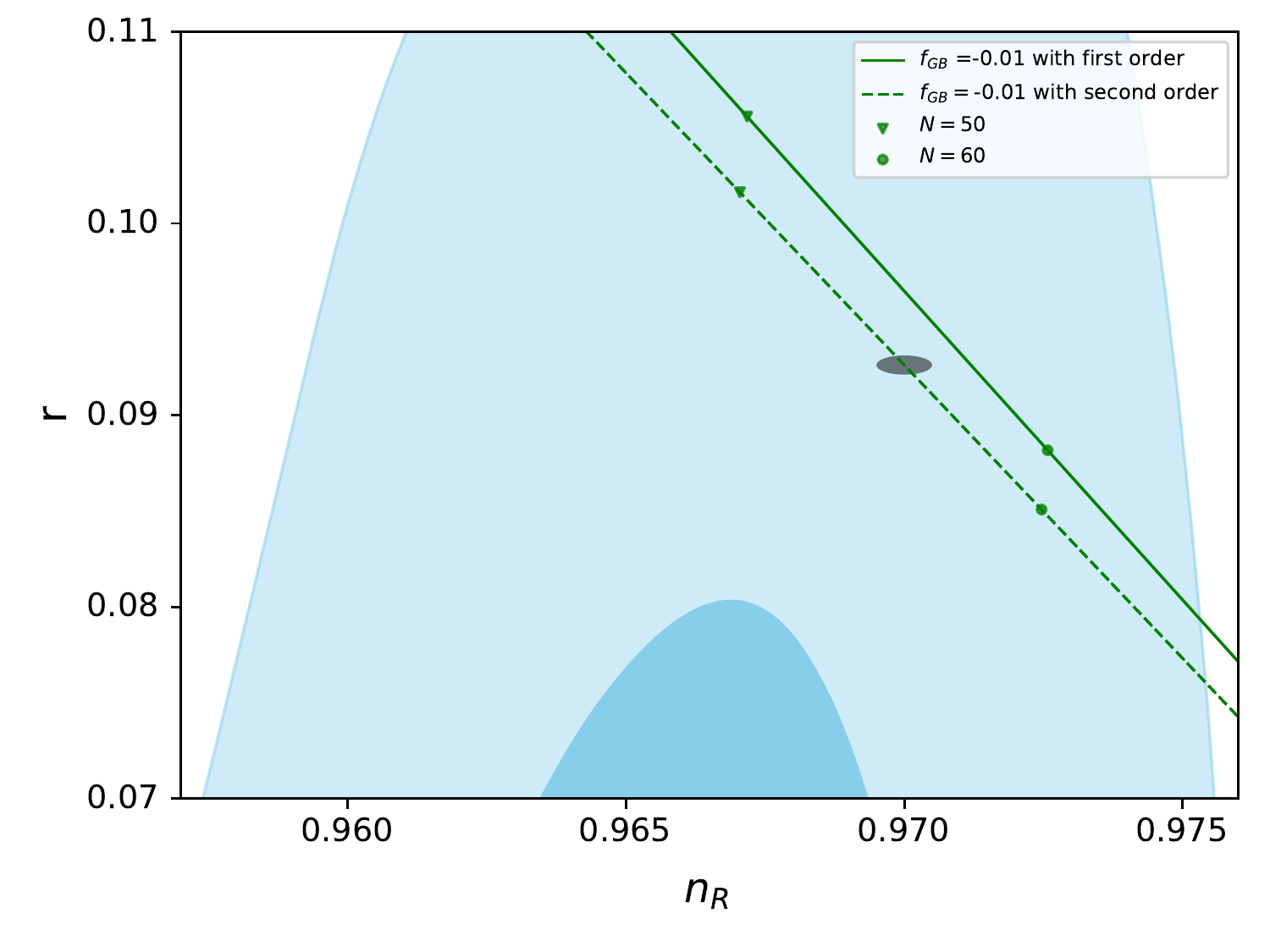}
\end{minipage}
}
\caption{{Difference between the $n_{\cal R}-r$ relations with the first- and the second-order corrections. The gray shaded $1-\sigma$ contour corresponds to a futuristic measurement with $\sigma_{n_{s-1}}=\sigma_r=10^{-3}$.}} 
\label{fig2}
\end{figure}

In Fig.~\ref{fig2}, the first-order and the second-order results of $n_{\cal R}$ and $r$ with Planck 2018 constraints in the models for different values of $N$ and $f_{\rm{GB}}$ are compared. The upper panel in Fig.~\ref{fig2} is for $f_{\rm{GB}}=0.02$, while the bottom panel is for $f_{\rm{GB}}=-0.01$, in which the shaded contour means a possible futuristic measurement with $\sigma(n_{\cal R}, r)<10^{-3}$. It is easy to find that for both $f_{\rm{GB}}=0.02$ and $f_{\rm{GB}}=-0.01$, the difference between the first-order and second-order predictions is larger than the futuristic experimental sensitivity on $n_{\cal R}$ and $r$, and this is even more obvious for $f_{\rm{GB}}=-0.01$ . This would make it convincing that the second-order corrections in the slow-roll approximation of $n_{\cal R}$ and $r$ in 4EGB inflation are necessary for fitting future experimental data.

\color{black}

\section{Conclusions and Outlook}

The uniform asymptotic approximation method is an error-controlled and systematically improvable method for constructing exact analytical solutions of linear perturbations. The effectiveness of this method has been verified in many applications, for example, in calculating primordial spectra for various inflation models \cite{Zhu:2013fha,Zhu:2013upa,Zhu:2014wfa,Zhu:2015ata, Zhu:2016srz,Zhu:2014wda,Zhu:2014aea,Zhu:2015xsa,Zhu:2015owa} and applications in studying the reheating process \cite{Zhu:2018smk} and quantum mechanics \cite{Li:2019cre}. 

In this paper, we applied the third-order uniform asymptotic approximation to derive the inflationary observables for scalar and tensor perturbations in the slow-roll inflation in the spatial covariant 4EGB gravity. With both scalar and tensor perturbations in terms of the flow of the Hubble flow slow-roll paramaters and the GB coupling constant, we obtained explicitly the analytical expressions of the primordial power spectra, spectral indices, running of spectral indices for both scalar and tensor perturbations, and the ratio between the tensor and scalar spectra up to the second-order in the expansions of the slow-roll approximation. Compared with the results obtained in the previous literature, the results presented in this paper represent the most accurate results obtained so far in the literature.

As we have mentioned in Sec.~I, the high-order corrections to the inflationary observables in the slow-roll expansions are very important in the future analysis of different inflation models with precise experimental data. For example, it is pointed out in \cite{Martin:2014rqa} that the future CMB experiments can measure both $n_{\mathcal{R}}$ and $r$ to be accurate with errors $\gtrsim 10^{-3}$. Thus, the contributions from the GB coupling constant to the scalar spectral index and the tensor-to-scalar ratio in general can not be ignored if they are at the magnitude of $\mathcal{O}(10^3)$. We expect the more precise forthcoming experimental data in the future can make significant constraint on the GB coupling constant in the spatial covariant 4EGB gravity, which could help us to understand the physics of the early universe.

\section*{Acknowledgements}

This work is supported in part by the National Key Research and Development Program of China Grant No.2020YFC2201503, and the Zhejiang Provincial Natural Science Foundation of China under Grant No. LR21A050001 and LY20A050002, the National Natural Science Foundation of China under Grant No. 12275238, No. 11975203, No. 11675143, and the Fundamental Research Funds for the Provincial Universities of Zhejiang in China under Grant No. RF-A2019015.

\appendix

\section*{Appendix A: General formulas of primordial spectra in the uniform asymptotic approximation}
\renewcommand{\theequation}{A.\arabic{equation}} \setcounter{equation}{0}

In this section, we present a brief introduction of the general formulas of primordial perturbations with a slow-varying sound speed and parameter $b(\tau)$ arsing from the nonlinear dispersion relation.

In the uniform asymptotic approximation, we first write Eqs.(\ref{1_scalar}) and (\ref{1_tensor}) in the standard form \cite{Zhu:2013upa}
\begin{eqnarray}
\frac{d^2 \mu(y)}{dy^2}=\{\lambda^2 \hat g(y)+q(y)\}\mu(y),\lb{standard}
\end{eqnarray}
where the new variables $y=-k\tau$, $\mu(y)=\mu_{\mathcal{R}}(y)$ and $\mu_h(y)$  correspond to scalar and tensor perturbations respectively, and
\begin{eqnarray}\lb{a2}
\lambda^2 \hat g(y)+q(y)=\frac{\nu^2(\tau)-1/4}{y^2}-c^2(\tau) - b(\tau) y^2.
\end{eqnarray}
where $c(\tau) = c_{s ,t}(\tau)$ is the effective sound speed for scalar and tensor perturbation modes respectively.  For scalar perturbation, we have \cite{Aoki:2020iwm}
\begin{eqnarray}\lb{cs}
\nu_{\mathcal{R}}^2(\tau) = \tau^2 \frac{z''(\tau)}{z(\tau)} + \frac{1}{4}\;\;{\rm with}\;\; c(\tau)=1\;\;  {\rm and}\;\; b(\tau)=0 ,\nb\\
\end{eqnarray}
and for tensor modes,
\begin{eqnarray}\lb{ct}
\nu_h^2(\tau) = \tau^2 \frac{a''(\tau)}{a(\tau)} + \frac{1}{4} .
\end{eqnarray}
It's worth noting that $\lambda$ is an assumed large parameter which is used to label the different approximate orders of the uniform asymptotic approximation, as we will see later in the square bracket of the general formulas of power spectrum in Eq.~(\ref{formula_pw}) in the square bracket. $\lambda=1$ can be set to $\lambda=1$ for simplicity in the ending calculation. Now we need to use the uniform asymptotic approximation to construct the approximate solutions of the above equation. In order to ensure the convergence of the errors of the approximate solutions, the only option is \cite{Zhu:2013upa}
\begin{eqnarray}
q(y)=- \frac{1}{4 y^2}.
\end{eqnarray}
 Then,  Eq.~(\ref{a2}) can be written as
\begin{eqnarray}\lb{lgy}
\lambda^2 \hat g(y)=\frac{\nu^2(\tau)}{y^2}-c^2(\tau) - b(\tau) y^2,
\end{eqnarray}
Here for scalar perturbation, $b(\tau) >0$ must meet the condition $b(\tau) >0$ for a healthy UV limit .

Since $b(\tau)$ is a small quantity, it is easy to find that the function $\lambda^2 \hat g(y)$ has a unique turning point, which is expressed as follows
\begin{eqnarray}
y_0^2(\bar \tau_0) &=&\frac{-c_0^2(\bar \tau_0) + \sqrt{\bar c_0^4(\tau_0) + 4 b(\bar \tau_0) \bar \nu_0^2(\bar \tau_0)}}{2 b(\bar \tau_0)}.
\end{eqnarray}
Then with reference to \cite{Zhu:2014wfa}, the general formula of the power spectrum can be written as follows
\begin{eqnarray}\lb{formula_pw}
\Delta^2(k) &\equiv& \frac{k^3}{4\pi^2} \left|\frac{u(y)}{z(\tau)}\right|^2_{y\to 0^+}\nb\\
&=&\frac{k^2}{8\pi^2}\frac{-k \tau}{z^2(\tau)\nu(\tau)}\exp{\left(2 \lambda  \int_y^{\bar y_0} \sqrt{\hat g(y')}dy' \right)}\nb\\
&&\times \left[1+\frac{\mathscr{H}(+\infty)}{\lambda}+\frac{\mathscr{H}^2(+\infty)}{2 \lambda^2}+\mathcal{O}\left(\frac{1}{\lambda^3}\right)\right].\nb\\
\end{eqnarray}
The order of the parameter $\lambda$ shows that the formula is in third order approximation. The integral of $\sqrt{g}$ in this formulas is given by Eqs.~(\ref{intg}) where $I_0$ and $I_1$ are given by Eqs.~(\ref{I0}) and (\ref{I1}), and the error control function is given by (\ref{error}).

Then we need to calculate the spectral indices. For this purpose, we primarily specify the $k$-dependence of $\bar\nu_0(\tau_0)$, $\bar \nu_1(\tau_0)$ through  $\tau_0 = \tau_0(k)$. From the relation $-k\tau_0= \bar y_0$, we obtain
\begin{eqnarray}
\frac{d\ln(-\tau_0)}{d\ln k} & =& -1 +\frac{d \ln \bar y_0}{d\ln (-\tau_0)}\frac{d \ln(-\tau_0)}{d \ln k},
\end{eqnarray}
which approximation is
\begin{eqnarray}
\frac{d\ln(-\tau_0)}{d\ln k} & \simeq &-1 -\frac{d \ln \bar y_0}{d\ln (-\tau_0)}.
\end{eqnarray}
Then with this relation, the spectral index can be written as
\begin{eqnarray}\lb{index}
n_{\mathcal{R}}-1, \; n_h &\equiv& \frac{d\ln \Delta^2_{s,t}}{d \ln k} \nb\\
&\simeq &3-2\bar{\nu}_0-\frac{2\bar{b}_1\bar{\nu}_0}{3\bar{c}_0^4}+\frac{2\bar{c}_1\bar{\nu}_0}{\bar{c}_0}+\frac{2\bar{b}_1\bar{\nu}_0^3}{3\bar{c}_0^4}\nb\\
&&-2\bar{\nu}_1\ln2+\frac{\bar{\nu}_1}{6\bar{\nu}_0^2}.
\end{eqnarray}
The running of the spectral index $\alpha \equiv dn/d\ln k$ can be similarly written as
\begin{eqnarray} \lb{running}
\alpha_{\mathcal{R},\; h}(k) &\simeq&2\bar{\nu}_1+\frac{2\bar{b}_2\bar{\nu}_0}{3\bar{c}_0^4}-\frac{2\bar{c}_2\bar{\nu}_0}{\bar{c}_0}-\frac{2\bar{b}_2\bar{\nu}_0^3}{3\bar{c}_0^4}\nb\\
&&+2\bar{\nu}_2\ln2-\frac{\bar{\nu}_2}{6\bar{\nu}_0^2}.
\end{eqnarray}

In the above, all the formulas (Eqs.(\ref{formula_pw}), (\ref{index}), and (\ref{running})) can be used to calculate the primordial perturbation spectra with different inflation models. It is easy to use these formulas, because they are determined only by the quantities $H(\tau)$, ($c_0, c_1, c_2$), ($\nu_0, \nu_1, \nu_2$) which are evaluated at the turning point. And these quantities can be easily obtain from Eqs.~(\ref{zs}, \ref{cs}) for scalar perturbations and Eqs.~(\ref{zt}, \ref{ct}) for tensor perturbations. In section IV, we apply these formulas to calculate the slow-roll power spectra for both scalar and tensor perturbations in the spatial covariant 4EGB gravity.

\section*{Appendix B: Integrals of $ \sqrt{g} $ and the error control function}
\renewcommand{\theequation}{B.\arabic{equation}} \setcounter{equation}{0}

From Eq.(\ref{lgy}) with $\lambda=1$, we have
\begin{eqnarray}\lb{gy}
g(y)=\frac{\nu^2(\tau)}{y^2}-c^2(\tau)-b(\tau)y^2.
\end{eqnarray}
where
\begin{eqnarray}\lb{v1}
\nu(\tau)\simeq\bar{\nu}_0+\bar{\nu}_1 \ln\frac{y}{\bar{y}_0}+\bar{\nu}_2\ln^2\frac{y}{\bar{y}_0},\\
c(\tau)\simeq\bar{c}_0+\bar{c}_1 \ln\frac{y}{\bar{y}_0}+\bar{c}_2\ln^2\frac{y}{\bar{y}_0},\lb{v2}\\
b(\tau)\simeq\bar{b}_0+\bar{b}_1 \ln\frac{y}{\bar{y}_0}+\bar{b}_2\ln^2\frac{y}{\bar{y}_0}.\lb{v3}
\end{eqnarray}
with
\begin{eqnarray}
\bar{\nu}_1 &\equiv &\frac{d\nu(\tau)}{d\ln(-\tau)}\Big|_{\tau_0},\bar{\nu}_2 \equiv\frac{d^2\nu(\tau)}{d\ln^2(-\tau)}\Big|_{\tau_0},\\
\bar{c}_1 &\equiv &\frac{dc(\tau)}{d\ln(-\tau)}\Big|_{\tau_0},\bar{c}_2 \equiv\frac{d^2c(\tau)}{d\ln^2(-\tau)}\Big|_{\tau_0},\\
\bar{b}_1 &\equiv &\frac{db(\tau)}{d\ln(-\tau)}\Big|_{\tau_0},\bar{b}_2 \equiv\frac{d^2b(\tau)}{d\ln^2(-\tau)}\Big|_{\tau_0}.
\end{eqnarray}
Substitute Eqs.(\ref{v1}), (\ref{v2}) and (\ref{v3}) into Eq.(\ref{gy}), we have 
\begin{eqnarray} 
\sqrt{g(y)}
=&&\frac{\sqrt{\bar{\nu}_0^2-\bar{c}_0^2y^2-\bar{b}_0y^4}}{y}\nb\\
&+&\frac{2\bar{\nu}_0\bar{\nu}_1-2\bar{c}_0\bar{c}_1y^2-b_1y^4}{2y\sqrt{\bar{\nu}_0^2-\bar{c}_0^2y^2-\bar{b}_0y^4}}\ln \frac{y}{\bar{y}_0}.
\end{eqnarray} 
Then we divided the integral $ \int \sqrt{g(y)}dy $ into two parts,
\begin{eqnarray}\lb{intg}
\int_y^{\bar{y}_0}\sqrt{g(y)}dy\simeq I_0+I_1,
\end{eqnarray}
where
\begin{eqnarray}\lb{I0} 
\lim_{y\rightarrow 0}I_0&\simeq&-\frac{\bar{\nu}_0}{2}-\frac{\bar{c}_0^2}{4\sqrt{\bar{b}_0}}\arctan\Big(\frac{2\sqrt{\bar{b}_0}\bar{\nu}_0}{\bar{c}_0^2}\Big)\nb\\
&&-\frac{1}{4}\bar{\nu}_0\ln(\bar{c}_0^4+4\bar{b}_0\bar{\nu}_0^2)-\bar{\nu}_0\ln\frac{y}{2\bar{\nu}_0},\\
\lim_{y\rightarrow 0}I_1&\simeq&\bar{b}_1\bar{\nu}_0^3\frac{5-6\ln {2}}{18\bar{c}_0^4}+\bar{\nu}_0\bar{c}_1\frac{1-\ln{2}}{\bar{c}_0}\nb\\
&&+\bar{\nu}_1\Big(-\frac{\pi^2}{24}+\frac{\ln^2 2}{2}-\frac{1}{2}\ln^2\frac{y}{\bar{y}_0}\Big).\lb{I1}
\end{eqnarray}
The integral form of error control function is
\begin{eqnarray}
\mathscr{H}(\xi)&\simeq&\frac{5}{36}\Big\{\int_{\bar{y}_0}^y\sqrt{g{y}dy}\Big\}^{-1}\Big|_{\bar{y}_0}^y\nb\\
&&-\int_{\bar{y}_0}^y\Big\{\frac{q}{g}-\frac{5{g'}^2}{16g^3}+\frac{g''}{4g^2}\Big\}\sqrt{g}dy.
\end{eqnarray} 
When $ y\rightarrow 0 $, the error control function can be written as
\begin{eqnarray} \lb{error}
\mathscr{H}(+\infty)&\simeq&\frac{\bar{c}_0^4+8\bar{b}_0\bar{\nu}_0^2}{6\bar{c}_0^4\bar{\nu}_0+24\bar{b}_0\bar{\nu}_0^3}\nb\\
&&+\frac{\bar{c}_1}{6\bar{c}_0\bar{\nu}_0}-\frac{23+12\ln{2}}{72\bar{\nu}_0^2}\bar{\nu}_1\nb\\
&&+\frac{(1+4\ln{2})\bar{\nu}_0}{6\bar{c}_0}\bar{b}_1.
\end{eqnarray}

\end{document}